\documentstyle[sprocl]{article}

\def\Journal#1#2#3#4{{#1} {\bf #2}, #3 (#4)}

\def\PLB{{\em Phys. Lett.}  B}

\def\PRD{{\em Phys. Rev.} D}
\def\PR{{\em Phys. Rev.}}

\def\be{\begin{equation}}
\def\ee{\end{equation}}
\def\bea{\begin{eqnarray}}
\def\eea{\end{eqnarray}}

\begin{document}


\title{ NO STRINGS ATTACHED\\
POTENTIAL vs. INTERACTION ENERGY IN QCD}
\author{T.\ Goldman}
\address{Theoretical Division, MSB283, Los Alamos National Laboratory,\\
Los Alamos, NM 87545\\E-mail: goldman@t5.lanl.gov}

\maketitle\abstracts{ 
In infrared-stable fixed-point field theories, the interaction energy
of a test particle is proportional to the non-relativistic (heavy
source) coordinate-space potential derived from the field strength
produced by that source. This is no longer true in ultraviolet-stable
fixed-point field theories (UVSFPFT) as they may not have a finite
infrared fixed point. This leads to the possibility that UVSFPFTs may
have quite conventional field strength distributions despite the
unusual spatial dependence expected for the interaction energy.}

\section{Introduction}

In infrared-stable fixed-point field theories, such as quantum
electrodynamics (QED), the quantum field theoretic renormalization
induced scale dependence of the effective coupling constant,
$\alpha(\mu^2) \equiv (g^2/4\pi)$, is innocuous at large distance
scales, because $\alpha(\mu^2)$ tends to a constant (1/137.036 for QED)
as $\mu$, the momentum scale at which the renormalization is defined,
tends to zero~\cite{mgmlr}.  As a result, the interaction energy ($V$)
between a test particle and a heavy (non-relativistic) source may be
described as the product of the coordinate-space potential produced by
the source multiplied by the charge `$g$' of the test particle. The
(static) potential, $\phi$, in turn is just the convolution of the
source charge distribution with the Green's function for the boson
which is the `force carrier'. It is therefore common to write $V$ in
the interaction Hamiltonian as $g \times \phi$ in a coordinate space
representation.

This orthodoxy has led to an unnecessary confusion in the extension of
such descriptions to the case of ultraviolet-stable fixed-point field
theories (UVSFPFTs), such as Quantum ChromoDynamics (QCD) which are
widely believed to lead to confining forces. In QCD, in particular, the
interaction Hamiltonian for a heavy source is frequently written as a
potential energy, and so, (incorrectly, I claim,) a (field strength)
potential which rises linearly with increasing modulus of the distance
from the source.  This is true despite more sophisticated momentum
space analyses~\cite{btgrr} which show that this is a property of the
interaction energy alone. The inexplicability of the spatial variation
of the field energy then leads to strange physical pictures wherein the
(field strength) potential of a heavy source is not spherically
symmetric, and indeed is undefined until a test particle is provided
which then defines a direction for a so-called `string' tying the heavy
source to the test particle. 

My purpose here is to point out that a much more conventional view of
the field distribution may still be tenable in such cases, provided one
retains the distinction between the field strength distribution (field
strength potential) of the force carrier and the interaction energy
(potential energy).

\section{Starting Point}
To this end, I employ the description of Goldhaber and Goldman~\cite{gg},
which identifies the force carrier of color confinement as a Lorentz
scalar (effective) boson, a composite of two or more gluons.
Ref.(\cite{gg}) begins with a perturbative approach to make the
discussion more specific.  In perturbation theory the coupling due to
gluon exchange between two quarks which are off mass shell by some
characteristic amount $\Delta M$ may be estimated at small four
momentum transfer squared $q^2$ by focusing on the most singular part
of the QCD coupling.

For exchange of the two-gluon color singlet combination, the
quark-quark potential in momentum space should be
\be
\tilde{V}(q^2) \approx \frac{(g_S(q^2))^{2}q^4}{(\Delta M)^2q^4},
\label{eq:1a}
\ee
where $(g_S(q^2))^2$ is $4\pi \alpha_{s}(q^2)$, the factor of $q^4$ in
the numerator comes from the integration over (small) loop momentum,
the $q^4$ in the denominator comes from the two gluon propagators, and
$(\Delta M)^2$ from the quark propagators.  We assume the correctness
of the Richardson~\cite{rich} ansatz for the leading behavior of
$\alpha_{s}$,
\be
\alpha_s(q^2) = \frac{12 \pi}{(33 - 2 n_{f}) ln(1 + q^2/\Lambda^{2})},
\label{eq:2}
\ee
where $\Lambda$ is of order the QCD scale but not necessarily equal to
$\Lambda_{\overline{MS}\,}$, and the coefficient is determined by the
one-loop $\beta$-function for QCD which depends on the number of light
quark flavors, $n_{f}$.  (For a recent confirmation that
$\alpha_s(q^2)$ at least diverges for $q^2 \rightarrow 0$, see
Ref.(\cite{hvsa}).) This implies a pole in $\alpha_{s}$ at $q^{2} = 0$,
giving a double pole in $\tilde{V}(q^2)$ and hence by Fourier
transformation a linearly rising potential $V(r)$ in coordinate space.

\section{Potential vs. Potential Energy}
My point, however, is that the result above is the interaction energy,
and cannot be simply translated into the field strength distribution.
The latter is found, in momentum space, from the product of the source
current, in this case, one of the two quarks, and the propagator for
the (effective) boson that is being exchanged. Since we are looking at
an exchange channel with $0^{+}$ quantum numbers for that (scalar)
boson, we first rewrite Eq.~\ref{eq:1a} in terms of that, as
\be
\tilde{V}(q^2) \approx \frac{(\alpha_{s}(q^2))^{2}}{(q^2 - m^2)},
\label{eq:1b}
\ee
where $m$ is the effective mass (at low $q^2$) for the scalar boson,
and we have ignored the normalization ($Z$) factor for its pole
strength, as the pole may not even actually exist. All that is needed
is that the propagator is approximately constant for $q^2 \approx 0$.
The potential of Eq.~\ref{eq:1b} has the same properties as
Eq.~\ref{eq:1a} if we identify $M$ with $m$, that is, as some sort of
minimal off-shellness required for the quarks for the whole picture to
apply.

Returning to the question of the field strength (potential, 
{\em not} potential energy), we find
\be
\tilde{\phi}(q^2) \approx \frac{(\alpha_{s}(q^2))}{(q^2 - m^2)},
\label{eq:3}
\ee
where $\phi$ is the field strength of the scalar boson. This has
entirely different properties from $V$ when Eq.~\ref{eq:2} is applied
as is appropriate for a UVSFPFT. There is now only a single pole (from
the one power of $\alpha_{s}$) at $q^{2} = 0$. Correspondingly, the
Fourier transform to coordinate space produces a potential $\phi(r)$
that varies only Coulombically at large $r$, that is
\be
\phi(r) \approx \frac{12 \pi \Lambda^{2}}{(33 - 2 n_{f}) m^2}
\frac{1}{r}.
\label{eq:4}
\ee
in the static limit. 

There is nothing exceptional about this field strength distribution:
The integral of the energy density associated with it is
logarithmically divergent in the infrared just as for the
electromagnetic Coulomb problem. There is a Gauss theorem and a
conserved, finite total charge enclosed, although the nature of that
charge is not apparent from these considerations. (See Ref.(\cite{gg})
for some conjectures on this point.) 

\section{Remaining Concerns}
Indeed, the only problem remaining is to understand why,
phenomenologically, there does not seem to be any van der Waals
potential associated with dipole fluctuations of color singlet systems.
This remains to be investigated, but some possibilities are immediately
apparent. One is the nature of the type of charge carried by the
sources~\cite{gg}. A second is the effective nature of the composite
boson -- below some minimal dipole separation, its coupling to the
charges may be significantly reduced. A third is that, again, the
fluctuations must be considered first in the field strength, not the
interaction energy, which is secondary from this point of view. The
rapid falloff with $r$ of conventional van der Waals ($r^{-7}$ when
retardation effects are taken into account) will thus lead to a more
rapid falloff with $r$ than in calculations which have not
distinguished between the potential and the potential energy (although
still not as rapid as the Yukawa potential, of course). Finally, the
energy excitation cost of the dipole formation further suppresses the
fluctuations which produce the van der Waals potential, and these are
much more severe in the case of a confining interaction energy than in
the conventional case.

\section{Conclusion}
I conclude that there is no well founded reason to reject the notion
that the potential produced by a (heavy) quark, or other isolated
(static) color source is of a Coulombic scalar form, while the
interaction energy of such a source with a test particle nonetheless
grows linearly with the separation between the source and the test
particle.
	
\section*{Acknowledgments}
This work was supported in part by the U.S. Department of  
Energy, Division of High Energy and Nuclear Physics, ER-23.

 \end{document}